\documentclass[preprint,preprintnumbers,amsmath,amssymb,color]{revtex4}

\usepackage{graphicx}
\usepackage{bm}

\begin{document}

\preprint{USTC-ICTS-11-06}

\title{~\\ \vspace{2cm}
Degrees of freedom of $f(T)$ gravity \vspace{1cm}}
\author{Miao Li$^{a}$}
\author{Rong-Xin Miao$^{b}$}
\author{Yan-Gang Miao$^{c}$}
\affiliation{$^{a}$Kavli Institute for Theoretical Physics, Key
Laboratory of Frontiers in Theoretical Physics, Institute of
Theoretical Physics, Chinese Academy of Sciences, \\
Beijing 100190, People's Republic of China. }\email{mli@itp.ac.cn}

\affiliation{$^{b}$Interdisciplinary Center for Theoretical Study, University of Science and Technology of China,\\
Hefei, Anhui 230026, People's Republic of
China.}\email{mrx11@mail.ustc.edu.cn}

\affiliation{$^{c}$School of Physics, Nankai University, Tianjin
300071, People's Republic of China.}\email{miaoyg@nankai.edu.cn}

\begin{abstract}
We investigate the Hamiltonian formulation of $f(T)$ gravity and
find that there are five degrees of freedom. The six first class
constraints corresponding to the local Lorentz transformation in
Teleparallel gravity become second class constraints in $f(T)$
gravity, which leads to the appearance of three extra degrees of
freedom and the violation of the local Lorentz invariance in $f(T)$
gravity. In general, there are $D-1$ extra degrees of freedom for
$f(T)$ gravity in $D$ dimensions, and this implies that the extra
degrees of freedom correspond to one massive vector field or one
massless vector field with one scalar field.
\end{abstract}



\maketitle

\section{Introduction}

 $f(T)$ gravity as an alternative to dark energy has recently received much attention in cosmology
 \cite{Ferraro1,Ferraro2,Bengochea:2009, Yu:2010a, Myrzakulov:2010a, Tsyba:2010,
 Linder:2010, Yi:2010b, Kazuharu:2010a, Kazuharu:2010b, Myrzakulov:2010b,
 Yu:2010c, Karami:2010, Dent:2010, Dent:2011,Huang,
 Sotiriou:2010,  Li, Li2, Tower, Cai,Ferraro3,Deliduman}.
 It is a generalization of the teleparallel gravity ($TG$) \cite{ein28,eineng,Aldrovandi} by replacing the so-called torsion
 scalar $T$ with $f(T)$. $TG$ was originally developed
 by Einstein in an attempt of unifying gravity and
 electromagnetism. The basic variables in $TG$ are tetrad fields $e_{a\mu}$, where the Weitzenbock connection rather than the Levi-Civita connection
 was used to define the covariant derivative. As a result, there is no
 curvature but only torsion.  A vector $V^\mu$ in $TG$ is parallel transported along a curve if its projection $V_a=e_{a\mu}V^\mu $ remains constant,
 this is the so-called teleparallelism. It is interesting that
 a covariant energy-momentum tensor of gravitation can naturally be
 defined in the gauge context of $TG$ \cite{Andrade}.

 $f(T)$ gravity is different from $f(R)$ gravity in several aspects.
 Firstly, as a main advantage compared with $f(R)$ gravity, the equations of
 motion of $f(T)$ gravity are second-order instead of fourth-order.
 Secondly, the local Lorentz invariance is violated in $f(T)$ gravity \cite{Li}. Therefore, extra degrees of
 freedom will appear. Till now, it is not clear how many extra
degrees of freedom
 there are in $f(T)$ gravity\footnote{Li et al. have studied the extra degrees of
freedom ($edof$) in $f(T)$ gravity  in light of the field
equations \cite{Li} and linear perturbation equations \cite{Li2} of $f(T)$
gravity. Their studies are very illuminating, however, the exact numbers of $edof$ are not
derived. From the method in Ref. \cite{Li}, the most natural
guess of the number of $edof$ is 6 since there are 6 more equations
in $f(T)$ gravity than that in $TG$. In the present paper, we obtain some new
insights on $edof$ of $f(T)$ gravity by investigating the constraint
structure. We find that the number of $edof$ is 3 instead of 6 and
the $edof$ seem to correspond to one massive vector field.}.
One might make a guess from the viewpoint of symmetry. There are 16
tetrad fields $e_{a\mu}$ in $f(T)$
 gravity among which four $e_{a0}$ are non-dynamical, so 12 degrees of freedom
 remain. Like $TG$, $f(T)$ gravity is invariant under the general coordinate
transformation which again removes four more degrees of freedom.
Because the local Lorentz invariance is violated in $f(T)$ gravity,
unlike $TG$, there is no further local gauge symmetries which can be
used to eliminate degrees of freedom. Therefore, one might guess
that there are totally eight (or six extra) degrees of freedom in
$f(T)$ gravity. However, it is not the case. The key point lies in
the fact that there are second class constraints in $f(T)$ gravity,
thus the above guess fails and the degrees of freedom should be
fewer than eight. To find out the number of degrees of freedom in
$f(T)$ gravity, we need to analyze the constraint structure
strictly. To the best of our knowledge, the Hamiltonian formulation  is our first choice to
derive the number of degrees of freedom\footnote{The main advantage of the Hamiltonian formulation is that one can
analyze constraint structures and then obtain the degrees of freedom ($dof$) directly by following Dirac's
procedure. On the other hand, in general, it is very hard to obtain
the independent dynamical $dof$ straight from the Lagrange or field
equations. One may try to find out $dof$ by analyzing the linear
perturbation equations of $f(T)$ gravity. However, for some special
backgrounds (for example, for the background
$e_{a\mu}={\rm diag}(1,1,1,1)$), a number of extra $dof$ do not appear in
the linear perturbation. Thus, the best way to derive
$dof$ is to adopt the Hamiltonian formulation.}. In this
paper, we analyze the constraint structure of $f(T)$ gravity and
find that there are totally five degrees of freedom. A simple
interpretation of this result is that the six first class
constraints corresponding to the local Lorentz transformation in
$TG$ turn into second class in $f(T)$ gravity, thus three extra
degrees of freedom emerge in $f(T)$ gravity.

 The paper is arranged as follows. In Sec.~2, we give a brief
 review of $f(T)$ gravity. In Sec.~3, we establish the Hamiltonian formulation of
 $f(T)$ gravity. In Sec.~4 and Sec.~5, we analyze the degrees of freedom of $f(T)$ gravity in $4D$ and $3D$,
 respectively. At the end of Sec.~5, we briefly discuss the the degrees of freedom of $f(T)$ gravity
 in $D$ dimensions. We conclude in Sec.~6.

\section{Brief review of $f(T)$ gravity}

Let us start with some definitions. $e_{a\mu}$ are tetrad fields and
$g_{\mu\nu}$ is the spacetime metric. They are related with each other by
$\eta_{ab}=e_{a\mu}e_{b\nu}g^{\mu\nu}=(-1,1,1,1)$ and
$g_{\mu\nu}=e_{a\mu}\eta^{ab}e_{b\nu}$, where $a$ and $\mu$ are the internal space and
spacetime indices, respectively. ``$a$" runs from 0 to 3, and $\mu=0, i$, ``$i$"
runs from 1 to 3. $T_{a\mu\nu}=\partial_\mu
e_{a\nu}-\partial_\nu e_{a\mu}$ are
torsion fields and $T$ is defined as
\begin{eqnarray}\label{T}
T=\Sigma^{abc}T_{abc},\qquad
\Sigma^{abc}=\frac{1}{4}(T^{abc}+T^{bac}-T^{cab})+\frac{1}{2}(\eta^{ac}T_d^{\
db}-\eta^{ab}T_d^{\ dc}).
\end{eqnarray}

 The Lagrangian density of $f(T)$ gravity is
\begin{eqnarray}\label{L}
L=- e f(T),
\end{eqnarray}
where we have set the Newton constant $G=\frac{1}{16\pi}$,
$e=|e_{a\mu}|=\sqrt{- g}$. In $TG$, $T$ is written as
\begin{eqnarray}\label{T}
T=-R-2\nabla^\mu T^\nu_{\ \ \mu\nu},
\end{eqnarray}
where $R$ and $\nabla^\mu$ are the Ricci scalar and covariant derivative
in Einstein gravity, respectively. Since $\nabla^\mu T^\nu_{\ \
\mu\nu}$ is
not a local Lorentz scalar, $f(T)$ gravity has no
local Lorentz invariance and thus it has more degrees of freedom than that of
$TG$ which is equivalent to Einstein gravity. One can
also argue that there are extra degrees of freedom for $f(T)$ gravity
by analyzing the equation of motion \cite{Li},
\begin{eqnarray}\label{motion}
H_{\mu\nu}=f'(T)(R_{\mu\nu}-\frac{R}{2}
g_{\mu\nu})+\frac{1}{2}g_{\mu\nu}[f(T)-f'(T)T]+2f''(T)\Sigma_{\nu\mu\rho}\nabla^\rho
T=\frac{1}{2}\Theta_{\mu\nu},
\end{eqnarray}
where $\Theta_{\mu\nu}$ is the stress-energy tensor of matter. For
simplicity, the action of matter field is supposed to have the local
Lorentz invariance, and therefore $\Theta_{\mu\nu}$ is symmetrical \cite{Li}.
There are six extra equations
\begin{eqnarray}\label{H[uv]}
H_{[\mu\nu]}=2f''(T)\Sigma_{[\nu\mu]\rho}\nabla^\rho T=0
\end{eqnarray}
in $f(T)$ gravity, thus it is expected that there are more physical
degrees of freedom. Naturally, one may guess that there are six
extra degrees of freedom, however, it is not the case. In fact, as we
will show in Sec. IV, there are only three extra degrees of freedom, which
implies that not all of the equations (eq.~(\ref{H[uv]})) contribute to the
dynamics of the tetrad fields.

To find out the number of the physical degrees of freedom, we shall
analyze the structure of constraints of $f(T)$ gravity in the next section. For
simplicity, let us rewrite the $f(T)$ Lagrangian density (eq.~(\ref{L})) in
an equivalent form
\begin{eqnarray}\label{L11}
L=- e [f(\varphi)+\phi(T-\varphi)],
\end{eqnarray}
where $\varphi$ and $\phi$ are two auxiliary fields. The variation of the action with respect to
$\varphi$ leads to the field equation
\begin{eqnarray}\label{E}
\phi=\frac{\partial f(\varphi)}{\partial\varphi}.
\end{eqnarray}
Using the above equation, we can solve $\varphi$ in terms of $\phi$
for every given function $f$. Substituting the solution
$\varphi(\phi)$ into eq.~(\ref{L11}), we can get an equivalent
Lagrangian density
\begin{eqnarray}\label{L2}
L=- e [\phi T+V(\phi)],
\end{eqnarray}
which contains only one auxiliary field $\phi$. We do not need to know the exact
form of $V(\phi)$ since it is irrelevant to our analysis of constraint
structure. We shall focus on the Lagrangian density eq.~(\ref{L2}) in the following sections.

To end this section, let us briefly discuss the conformal rescaling
of the action of $f(T)$ gravity. It is well known that the action of
f(R) gravity
\begin{equation}\label{Sf}
 S=\int d^4x\sqrt{-g}f(R)
\end{equation}
is equivalent to the Einstein gravity with a scalar field
\begin{equation}\label{CSf}
S=\int
d^4x\sqrt{-\tilde{g}}[\tilde{R}-\frac{1}{2}\partial^\mu\tilde{\phi}\partial_\mu\tilde{\phi}-V(\tilde{\phi})],
\end{equation}
if we perform conformal transformation $\tilde{g}_{ab}=f'g_{ab}$ and
$\tilde{\phi}=\sqrt{3}\ln f'$,
$V(\tilde{\phi})=\frac{Rf'-f}{(f')^2}$. Similarly, performing the
conformal transformation $\tilde{e}_{a\mu}=\sqrt{\phi}e_{a\mu}$ and
$\tilde{\phi}=\sqrt{3}\ln \phi$, $V(\tilde{\phi})=\frac{V(\phi)
}{\phi^2}$, we can rewrite the action eq.~(\ref{L2}) in the
following form
\begin{equation}\label{L3}
S=\int
d^4x\sqrt{-\tilde{g}}[\tilde{R}+\frac{1}{2}\partial^\mu\tilde{\phi}\partial_\mu\tilde{\phi}-V(\tilde{\phi})-\frac{2}{\sqrt{3}}\tilde{T}^{\nu\mu}_{\
\ \ \nu}\partial_\mu \tilde{\phi} ],
\end{equation}
where we have used formulas $e=\sqrt{-g}$ and $T=-R-2\nabla^\mu
T^\nu_{\ \ \mu\nu}$. Note that the kinetic energy term of scalar
$\tilde{\phi}$ in the above action has a wrong sign which seems to
lead to instabilities. The last term in the above action is not a
local Lorentz scalar which is a reflection of the violation of the
local Lorentz invariance in f(T) gravity.

The above action may realize a kind of Higgs mechanism, so that a
vector in $e_{a\mu}$ becomes a massive dynamic vector. Thus, it is
possible that the general $f(T)$ gravity contains a massless spin
two graviton and a massive vector.

\section{Hamiltonian formulation of $f(T)$ gravity}

Following the procedure developed in refs.~\cite{Maluf1,Maluf2}, we analyze the
Hamiltonian formulation of $f(T)$ gravity in this section. From the
Lagrangian density eq.(\ref{L2}), we can derive the momenta
conjugate to $e_{a\mu}$ and $\phi$, respectively,
\begin{eqnarray}\label{momentum}
& &\Pi^{a\mu}=\frac{\partial L}{\partial(\partial_0 e_{a\mu})}=-4\phi e \Sigma^{a0\mu}, \\
& &\pi=\frac{\partial L}{\partial(\partial_0 \phi)}=0.
\end{eqnarray}
There are eleven primary constraints in the above equations
\begin{eqnarray}\label{constraint}
\Gamma^{ab}&=&\Pi^{ab}-\Pi^{ba}+2\phi e\left[e^{am}e^{bj}T^0_{\ mj}-(e^{am}e^{b0}-e^{bm}e^{a0})T^j_{\ mj}\right]\approx 0, \\
\Pi^{a0}&\approx& 0, \label{con1} \\
\pi &\approx& 0. \label{con2}
\end{eqnarray}
The derivation of constraints eq.~(\ref{constraint}) is very
complicated, please refer to ref.~\cite{Maluf1} for details.
Constraints eqs.~(\ref{con1}) and (\ref{con2}) are obvious. Since
the Lagrangian density eq.~(\ref{L2}) contains no time derivatives of
$e_{a0}$ and $\phi$, their conjugate momenta
eqs.~(\ref{con1}) and (\ref{con2}) vanish. We can derive the primary
Hamiltonian density by a similar method to that given in
ref.~\cite{Maluf1}. Here we only need to replace $k$ appearing in ref.~\cite{Maluf1} by $\phi$ and
add a term of potential energy, we give the result below:
\begin{eqnarray}\label{H0}
H_0&=&\Pi^{a\mu}\dot e_{a\mu} + \pi\dot \phi -L \nonumber\\
   &=&-e_{a0}\partial_k \Pi^{ak} -{1\over
{4g^{00}}} \phi e \biggl(g_{ik}g_{jl}P^{ij}P^{kl}- {1\over
2}P^2\biggr)\nonumber\\
& &+\phi e\biggl( {1\over 4}g^{im}g^{nj}T^a\,_{mn}T_{aij} +{1\over
2}g^{nj}T^i\,_{mn}T^m\,_{ij} -g^{ik}T^j\,_{ji}T^n\,_{nk}\biggr)+e
V(\phi),
\end{eqnarray}
where $P^{ik}=\frac{1}{\phi e}\Pi^{(ik)}-\Delta^{ik}$,
$P=P^{ij}g_{ij}$, and
\begin{eqnarray}\label{D}
\Delta^{ik}\;=\;-g^{0m}(
g^{kj}T^i\,_{mj}+g^{ij}T^k\,_{mj}-2g^{ik}T^j\,_{mj})
-(g^{km}g^{0i}+g^{im}g^{0k}) T^j\,_{mj}.
\end{eqnarray}
The total Hamiltonian density is given by
\begin{eqnarray}\label{H}
H=H_0+\lambda_{ab}\Gamma^{ab}+\lambda \pi,
\end{eqnarray}
where $\lambda_{ab}$ and $\lambda$ are Lagrange multipliers. We have
ignored the Lagrange multipliers corresponding to $\Pi^{a0}$ because
$\Pi^{a0}$ are first class constraints in view of the non-dynamical
character of tetrad fields $e_{a0}$ ( We have checked that adding
Lagrange multipliers corresponding to $\Pi^{a0}$ do not affect the
conclusions of this paper ).

The basic Poisson brackets of the canonical variables are
\begin{eqnarray}\label{Poisson}
& &\{ e_{a\mu}(x), \Pi^{b\nu}(y)\}=\delta^b_a \delta^\nu_\mu
\delta^3(x-y),\nonumber\\
& &\{ \phi(x), \pi(y) \}=\delta^3(x-y),
\end{eqnarray}
with which we can calculate the Poisson brackets of two arbitrary
fields.

Now let us begin to search for secondary constraints. From
$\{\Pi^{a0}, H \}\approx 0 $, we can derive four secondary
constraints
\begin{eqnarray}\label{Ca}
C^a\;&=&\;-\partial_k \Pi^{ak}+ ee^{a0}V(\phi)+
e^{a0}\biggl[-{1\over{4g^{00}}}\phi
e\biggl(g_{ik}g_{jl}P^{ij}P^{kl}-
{1\over 2}P^2\biggr)\nonumber\\
& &+\phi e\biggl( {1\over 4}g^{im}g^{nj}T^b\,_{mn}T_{bij}+ {1\over
2}g^{nj}T^i\,_{mn}T^m\,_{ij} -g^{ik}T^m\,_{mi}T^n\,_{nk}\biggr)
\biggr]\nonumber\\
& &-{1\over {2g^{00}}}\phi e\biggl(g_{ik}g_{jl}\gamma^{aij}P^{kl}-
{1\over 2}g_{ij}\gamma^{aij}\,P\biggr)
-\phi e\,e^{ai}\biggl(g^{0m}g^{nj}T^b\,_{ij}T_{bmn}\nonumber\\
& &+g^{nj}T^0\,_{mn}T^m\,_{ij}+g^{0j}T^n\,_{mj}T^m\,_{ni}
-2g^{0k}T^m\,_{mk}T^n\,_{ni}-2g^{jk}T^0\,_{ij}T^n\,_{nk} \biggr)\;
\end{eqnarray}
where $\gamma^{aij}$ is defined by
\begin{eqnarray}
\gamma^{aij}\;&=&\;-{1\over {2ke}}(e^{ai}e^{b0}e^{cj}\Gamma_{bc}+
e^{aj}e^{b0}e^{ci}\Gamma_{bc}) -e^{ak}\biggl[ g^{00}(
g^{jm}T^i\,_{km}+g^{im}T^j\,_{km}+
2g^{ij}T^m\,_{mk})\nonumber\\
&+&g^{0m}(g^{0j}T^i\,_{mk}+g^{0i}T^j\,_{mk})
-2g^{0i}g^{0j}T^m\,_{mk}
+(g^{jm}g^{0i}+g^{im}g^{0j}-2g^{ij}g^{0m})T^0\,_{mk} \biggr]\;.\nonumber\\
\end{eqnarray}
It is interesting that $C^a$ can be written in the following form
\begin{eqnarray}\label{Ca1}
& &C^a=e^{a0}H_0 + e^{ai}H_i,\\
& &H_i=-e_{bi}\partial_k \Pi^{bk}-\Pi^{bk} T_{bki}.
\end{eqnarray}

For $TG$, there are no further secondary constraints, and all the
constraints $\Gamma^{ab}$, $H_0$, $H_i$, and $\Pi^{a0}$ are first
class~\cite{Maluf1}. $\Gamma^{ab}$ are the generators of six local
Lorentz transformations, and $H_0, H_i$ are that of four general
coordinates transformations. $\Pi^{a0}$ can be used to fix the
tetrad fields $e_{a0}$, which is consistent with the fact that
$e_{a0}$ are not dynamical fields. Thus, the physical degrees of
freedom of $TG$ are $\frac{2n-2m-l}{2}=2$, where ``$n=16$" is the
number of fields, ``$m=14$" is the number of first class
constraints, and ``$l=0$" is the number of second class constraints.

We note that the situation is very different for $f(T)$ gravity. Poisson
brackets between $\Gamma^{ab}$,  $H_0$, and $\pi$ no longer vanish because
$\phi$ is now a function rather than a constant, which is different from the case
of $TG$. We shall make a careful analysis for $f(T)$ gravity in the next
section.

\section{Degrees of freedom of $f(T)$ gravity in $4D$}
We calculate the Poisson brackets among $\Gamma^{ab}$,
$\Pi^{a0}$, $H_0$, $H_i$, and $\pi$, and give one very complicated
secondary constraint for $f(T)$ gravity in $4D$. We analyze the
structure of constraints and find that there are five degrees of
freedom in all.

Since $\Pi^{a0}$ and $H_i$ are independent of $\phi$ and $\pi$,
similar to ref.~\cite{Maluf1}, Poisson brackets between $\Pi^{a0}$,
together with $H_i$, and the other constraints still vanish.  The
other non-vanishing Poisson brackets are listed below:
\begin{eqnarray}
\{\Gamma^{ab}(x),\Gamma^{cd}(y)\}&\approx&\left[-\eta^{ac}G^{bd}-\eta^{bd}G^{ac}+\eta^{bc}G^{ad}+\eta^{ad}G^{bc}\right]\delta^3(x-y),
\label{constraints1} \\
\{\Gamma^{ab}(x),\pi(y)\}&\approx& 2 e\left[e^{am}e^{bj}T^0_{\
mj}-(e^{am}e^{b0}-e^{bm}e^{a0})T^j_{\ mj}\right] \delta^3(x-y),\label{constraints2} \\
\{H_0(x), \Gamma^{ab}(y)\}&\approx &
-\Bigg[\frac{e}{g^{00}}(g_{ik}g_{jl}-\frac{1}{2}g_{ij}g_{kl})P^{kl}\Big(e^{am}e^{bj}g^{0i}-e^{bm}e^{aj}g^{0i} \nonumber \\
& & -(e^{am}e^{b0}-e^{bm}e^{a0})g^{ij}+(e^{aj}e^{bo}-e^{bj}e^{a0})g^{im}\Big) \nonumber \\
& & +2e\Big(e^{an}e^{bj}T^m_{\ nj}-(e^{an}e^{bm}-e^{bn}e^{am})T^j_{\ nj}\Big)\Bigg]\partial_m \phi
\delta^3(x-y),\label{constraints3} \\
\{\pi(x), H_0(y)\}& \approx & \Bigg[{1\over {4g^{00}}} e
\biggl(g_{ik}g_{jl}P^{ij}P^{kl}- {1\over 2}P^2\biggr)-\frac{1}{2\phi
g^{00}}(g_{ik}g_{jl}-\frac{1}{2}g_{ij}g_{kl})P^{kl}\Pi^{ij}-e
\frac{\partial V(\phi)}{\partial \phi} \nonumber \\
& & -e\biggl( {1\over 4}g^{im}g^{nj}T^a\,_{mn}T_{aij} -{1\over
2}g^{nj}T^i\,_{mn}T^m\,_{ij}
-g^{ik}T^j\,_{ji}T^n\,_{nk}\biggr)\Bigg]\
\delta^3(x-y),\label{constraints4}\nonumber\\
\end{eqnarray}
where $G^{bd}=2e(e^{bm}e^{d0}-e^{dm}e^{b0})\partial_m \phi$ in eq.~(\ref{constraints1}), and `$\approx$' denotes the Dirac's weak equality.
Eqs.~(\ref{constraints1})-(\ref{constraints3}) imply that the local Lorentz invariance is violated.

Now let us begin to search for other secondary constraints. The
consistency of constraints $H_0$, $\Gamma^{ab}$ and $\pi$ requires
\begin{eqnarray}\label{equ}
\  \{ H_0, H \}&=&\{ H_0, H_0 \}+\{ H_0, \Gamma^{cd} \}\lambda_{cd}+\{ H_0, \pi \}\lambda\approx 0, \nonumber\\
\{ \Gamma^{ab}, H \}&=&\{ \Gamma^{ab}, H_0 \}+\{ \Gamma^{ab}, \Gamma^{cd} \}\lambda_{cd}+\{ \Gamma^{ab}, \pi \}\lambda \approx 0,\nonumber\\
\ \{ \pi, H \}&=&\{ \pi, H_0 \}+\{ \pi, \Gamma^{cd} \}\lambda_{cd}
\approx 0.
\end{eqnarray}
We want to mention that the second equation above is equivalent to
the field equation eq.~(\ref{H[uv]}), and the third equation is
equivalent to $\frac{\partial V(\phi)}{\partial \phi}+T=0$. There
are eight equations but only seven unknown quantities
$\lambda_{cd}$ and $\lambda$, thus it is expected that we can derive
one secondary constraint from the above equations. It is indeed the
case. For simplicity, we use $\Gamma^i=e_a^{\ 0}e_b^{\
i}\Gamma^{ab}$ and $\Gamma^{ij}=e_a^{\ i}e_b^{\ j}\Gamma^{ab}$
instead of $\Gamma^{ab}$ in the following calculations, and note
that $\{ \Gamma^i, \Gamma^j \}\approx 0$, $\{ H_0, \Gamma^i
\}\approx \Pi^{(im)}\partial_m \ln(\phi)$. Define
\begin{eqnarray}\label{y}
y_i=\{ H_0, \Gamma^i \},\qquad y_4=\{ H_0, \Gamma^{12} \},\qquad y_5=\{
H_0, \Gamma^{13} \},\qquad y_6=\{ H_0, \Gamma^{23} \},
\end{eqnarray}
\begin{eqnarray}\label{x}
x_0= \{ H_0, \pi \},\qquad x_i=\{\Gamma^i, \pi \},\qquad
x_4=\{\Gamma^{12}, \pi \},\qquad x_5=\{\Gamma^{13}, \pi \},\qquad
x_6=\{\Gamma^{23}, \pi \},\nonumber\\
\end{eqnarray}
\begin{eqnarray}\label{A}
A_{i1}&=&\{\Gamma^i, \Gamma^{12} \}\nonumber\\
 &\approx& 2e\left[g^{0i}(g^{01}g^{2m} - g^{02}g^{1m}) + g^{1i} (g^{0m}g^{02} - g^{2m}g^{00})
- g^{i2}(g^{0m}g^{01} - g^{1m}g^{00})\right]\partial_m \phi, \nonumber\\
A_{i2}&=&\{\Gamma^i, \Gamma^{13} \}\nonumber\\
 &\approx& 2e\left[g^{0i}(g^{01}g^{3m} - g^{03}g^{1m}) + g^{1i} (g^{0m}g^{03} - g^{3m}g^{00})
- g^{i3}(g^{0m}g^{01} - g^{1m}g^{00})\right]\partial_m \phi, \nonumber\\
A_{i3}&=&\{\Gamma^i, \Gamma^{23} \}\nonumber\\
 &\approx& 2e\left[g^{0i}(g^{02}g^{3m} - g^{03}g^{2m}) + g^{2i} (g^{0m}g^{03} - g^{3m}g^{00})
- g^{i3}(g^{0m}g^{02} - g^{2m}g^{00})\right]\partial_m \phi,\nonumber\\
\end{eqnarray}
\begin{eqnarray}\label{B}
B_{12}&=&\{\Gamma^{12}, \Gamma^{13} \}\nonumber\\
 &\approx& 2e\left[g^{12}(g^{1m}g^{03} - g^{3m}g^{01})-g^{11}(g^{2m}g^{03}-g^{3m}g^{02})+
 g^{13} (g^{2m}g^{01} - g^{1m}g^{02})\right]\partial_m \phi, \nonumber\\
B_{13}&=&\{\Gamma^{12}, \Gamma^{23} \}\nonumber\\
 &\approx& 2e\left[g^{22}(g^{1m}g^{03}- g^{3m}g^{01})-g^{12}(g^{2m}g^{03} - g^{3m}g^{02})-g^{23}(g^{1m}g^{02}-g^{2m}g^{01})\right]\partial_m \phi,
 \nonumber\\
B_{23}&=&\{\Gamma^{13}, \Gamma^{23} \}\nonumber\\
 &\approx& 2e\left[g^{23}(g^{1m}g^{03}- g^{3m}g^{01})-g^{33}(g^{1m}g^{02}-g^{2m}g^{01})+
 g^{13}(g^{3m}g^{02} - g^{2m}g^{03})\right]\partial_m \phi,\nonumber\\
\end{eqnarray}
we can rewrite Eq.~(\ref{equ}) in a compact form
\begin{equation}\label{MLamma}
M \Lambda =0,
\end{equation}
where $\Lambda=(1, \lambda_1, \lambda_2, \lambda_3,
\lambda_4,\lambda_5, \lambda_6, \lambda)^T$,
$\lambda_i=e_{a0}e_{bi}\Gamma^{ab},\
\lambda_4=e_{a1}e_{b2}\Gamma^{ab},\
\lambda_5=e_{a1}e_{b3}\Gamma^{ab},\
\lambda_6=e_{a2}e_{b3}\Gamma^{ab}$ and matrix $M$ is given by
\begin{equation}
M=\left(
\begin{array}{cccccccc}
 0 & y_1 & y_2 & y_3 & y_4 & y_5 & y_6 & x_0\\
 -y_1 & 0 & 0 & 0 & A_{11} & A_{12} & A_{13} & x_1 \\
 -y_2 & 0 & 0 & 0 & A_{21} & A_{22} & A_{23} & x_2 \\
 -y_3 & 0 & 0 & 0 & A_{31} & A_{32} & A_{33} & x_3 \\
 -y_4 & -A_{11} & -A_{21} & -A_{32} & 0 & B_{12} & B_{13} & x_4\\
 -y_5 & -A_{12} & -A_{22} & -A_{32} & -B_{12} & 0 & B_{23} & x_5\\
 -y_6 & -A_{13} & -A_{23} & -A_{33} & -B_{13} & -B_{23}  & 0  & x_6\\
 -x_0 & -x_1 & -x_2 & -x_3 & -x_4 & -x_5 & -x_6 & 0
\end{array}
\right).
\end{equation}
Because Eq.~(\ref{MLamma}) has a nonzero solution $\Lambda=(1,
\lambda_1, \lambda_2, \lambda_3, \lambda_4,\lambda_5, \lambda_6,
\lambda)^T$, the determinant of matrix $M$ should vanish. Thus, we
get one constraint $\mid M \mid \approx 0$. It is too complicated to
use $\mid M \mid$ as a secondary constraint, therefore we try to
simplify it. Note that $M$ is an $8\times8$ antisymmetric matrix
whose determinant can be written as $D^2$ where $D$ is a function of
the elements of $M$. Applying eqs.~(\ref{A}) and (\ref{B}), we find
that $x_0$ does not contribute to $\mid M \mid$. Now, we can express
the secondary constraint as
\begin{eqnarray}\label{pi}
\pi_1&=&\sqrt{\mid M \mid}\ \mid_{x_0=0}\nonumber\\
&=&y_1\biggl[(A_{33}B_{12}- A_{32}B_{13}+ A_{31}B_{23})x_2-(A_{23}B_{12}-A_{22}B_{13}+A_{21}B_{23})x_3 \nonumber\\
& &+(A_{23}A_{32}-A_{22}A_{33})x_4-(A_{23}A_{31}-A_{21}A_{33})x_5 +(A_{22}A_{31}-A_{21}A_{32})x_6\biggr]\nonumber \\
& &+y_2\biggl[-(A_{33} B_{12}- A_{32} B_{13} + A_{31} B_{23} )x_1 + (A_{13} B_{12}- A_{12} B_{13}+
 A_{11} B_{23} )x_3\nonumber\\
& & -(A_{13} A_{32} - A_{12} A_{33} )x_4 + (A_{13} A_{31}- A_{11} A_{33} )x_5 -
 (A_{12} A_{31} -A_{11} A_{32}) x_6\biggr]\nonumber\\
& &+y_3\biggl[(A_{23} B_{12}- A_{22} B_{13}+ A_{21} B_{23})x_1 - (A_{13} B_{12}- A_{12} B_{13}+ A_{11}
B_{23} x_2)\nonumber\\
& &+(A_{13} A_{22}- A_{12} A_{23} )x_4 - (A_{13} A_{21}- A_{11} A_{23} )x_5 +(A_{12} A_{21} - A_{11} A_{22} )x_6 \biggr]\nonumber\\
& &+y_4\biggl[ -(A_{23} A_{32}-A_{22} A_{33}) x_1 + (A_{13} A_{32}- A_{12} A_{33} )x_2 - (A_{13}
A_{22} - A_{12} A_{23}) x_3\biggr]\nonumber\\
& &+y_5\biggl[ (A_{23} A_{31}- A_{21} A_{33} )x_1 - (A_{13} A_{31}- A_{11} A_{33}) x_2 +( A_{13}
A_{21}-A_{11} A_{23} )x_3\biggr]\nonumber\\
& &+y_6\biggl[ -(A_{22} A_{31}- A_{21} A_{32} )x_1 + (A_{12} A_{31}-
A_{11} A_{32} )x_2 - (A_{12} A_{21} -A_{11} A_{22} )x_3\biggr],\nonumber\\
\end{eqnarray}
which is a very complicated formula. In general, the constraint $\pi_1$ takes the form
$C^{(mnl)}\partial_m\phi\partial_n\phi\partial_l\phi=0$, where
$C^{(mnl)}$ is independent of the space derivatives of $\Pi^{ai},\phi,
T_{aij}$. If the metric $g_{\mu\nu}$ has only diagonal elements,
$\pi_1$ can  be highly simplified as
\begin{eqnarray}\label{pi1}
(\Pi^{(mn)}g^{il}g^{jh}g^{kg}T^0_{\
hg}-\Pi^{0m}g^{in}g^{jh}g^{kg}T^l_{\
hg})\epsilon_{ijk}\partial_m\phi\partial_n\phi\partial_l\phi=0.
\end{eqnarray}

 Since $M$ is an $8\times8$ antisymmetric matrix with zero determinant,
 the rank of $M$ is 6. Thus, after imposing $\pi_1=0$, there are only
 six independent equations in eq.~(\ref{MLamma}) for seven Lagrange multipliers. The
 consistency condition of constraint $\pi_1$
\begin{eqnarray}\label{C1}
\{ \pi_1, H \}&=&\{ \pi_1, H_0 \}+\{ \pi_1, \Gamma^{cd}
\}\lambda_{cd} +\{ \pi_1, \pi \}\lambda \approx 0
\end{eqnarray}
leads to another equation for the Lagrange multipliers. We shall prove in the
appendix that eq.~(\ref{C1}) together with eq.~(\ref{MLamma})
provides seven independent equations for the seven Lagrange multipliers
$\lambda_{ab}$ and $\lambda$. As a result, all the Lagrange
multipliers can be determined and there are no further secondary
constraints.

We turn to analyze the structure of constraints. In order to find
out the number of degrees of freedom in $f(T)$ gravity, we do not
need to calculate the Poisson brackets between $\pi_1$ and the other
constraints $\Gamma^{ab}$, $\pi$, $\Pi^{a0}$, $H_0$, $H_i$ because the
calculation is highly complicated due to eq.~(\ref{pi}). With the
results at hand, we are ready to give the number of degrees of
freedom. Let us recall that the Poisson brackets between ($\Pi^{a0}
, H_i, H$) and ($\Gamma^{ab}, \pi $) are zero. Consequently, the
Poisson brackets among all the constraints ($\pi_1, \pi,
\Gamma^{ab}, \Pi^{a0}, H_i, H$) take the following form
$$
\begin{array}{c@{\hspace{-5pt}}l}
N=\left(\begin{array}{ccc|ccc}
0&\{\pi_1, \pi \} & \{\pi_1, \Gamma^{ab} \}&\{\pi_1, \Pi^{a0} \} &\{\pi_1, H_i \} &0\\
\{\pi , \pi_1 \} & 0 & \{\pi, \Gamma^{ab} \} & 0 & 0 & \ \ 0 \\
 \{\Gamma^{cd}, \pi_1\} & \{\Gamma^{cd}, \pi \} & \{\Gamma^{cd}, \Gamma^{ab}\} & 0 & 0 & 0\\\hline
 \{\Pi^{a0}, \pi_1\} & 0 & 0 & 0 &0 &0 \\
 \{H_i, \pi_1\} & 0 & 0 & 0 & 0 & 0\\
 0 & 0 & 0 & 0 & 0 & 0
\end{array}\right).
&\begin{array}{l}\left.\rule{0mm}{8mm}\right\}8\\
\\\left.\rule{0mm}{8mm}\right\}8
\end{array}\\[-5pt]
\begin{array}{cc}\underbrace{\rule{50mm}{0mm}}_8&\
\underbrace{\rule{30mm}{0mm}}_8\end{array}&
\end{array}
$$
This is a $16\times 16$ antisymmetric matrix. The top left corner of
it must be an $8\times 8$ nonsingular matrix, otherwise, we would
not solve all of the Lagrange multipliers. The lower right corner is
an $8\times 8$ zero matrix. The lower left quarter, denoted by
$N_{ll}$, is an $8\times 8$ matrix with the non-vanishing first line
only. Obviously, the rank of $N_{ll}$ is at most 1, we can turn it
into a matrix with only the non-vanishing element, denoted as
$N_{ll(11)}$, by applying the elementary transformations of
matrices. In addition, we can make a similar treatment to the top
right corner of $N$. Therefore, the non-zero part of $N$ becomes a
$9\times 9$ antisymmetric matrix in the top left corner whose rank
is eight, which will be shown below. Now it is clear that the rank
of $N$ is eight.
$$
\begin{array}{c@{\hspace{-5pt}}l}
\left(\begin{array}{cccc|cc}
0&\{\pi_1, \pi \} & \{\pi_1, \Gamma^{ab} \}&-N_{ll(11)} &0 &0\\
\{\pi , \pi_1 \} & 0 & \{\pi, \Gamma^{ab} \} & 0 & 0 & 0 \\
 \{\Gamma^{cd}, \pi_1\} & \{\Gamma^{cd}, \pi \} & \{\Gamma^{cd}, \Gamma^{ab}\} & 0 & 0 & 0\\
 N_{ll(11)} & 0 & 0 & 0 &0 &0 \\\hline
0 & 0 & 0 & 0 & 0 & 0\\
 0 & 0 & 0 & 0 & 0 & 0
\end{array}\right).
\end{array}
$$

The above discussions show that there are eight second class constraints
$(\Gamma^{ab}, \pi, \pi_1)$ together with eight first class constraints,
thus the degrees of freedom of $f(T)$ gravity are
$\frac{2n-2m-l}{2}=5$, where ``$n=17$" is the number of fields,
``$m=8$" is the number of first class constraints and ``$l=8$" is
the number of second class constraints. This conclusion is
consistent with the physical analysis of $f(T)$ gravity: The action
of $f(T)$ gravity is invariable under the general coordinate
transformation, and $e_{a0}$ are not dynamical fields. Similar to
$TG$, four of the eight first class constraints correspond to the
general coordinate transformation of $f(T)$ gravity; the rest four
first class constraints can be used again to fix the non-dynamical
fields $e_{a0}$. Two of the second class constraints $(\pi, \pi_1)$
can be used to eliminate the auxiliary field $\phi$ introduced in
the action eq.~(\ref{L2}), and the existence of the other six second
class constraints ($\Gamma^{ab}$) implies that the local Lorentz
invariance is violated completely.

\section{Degrees of freedom of $f(T)$ gravity in $3D$}

In this section, we establish the Hamiltonian formulation of $f(T)$
gravity in $3D$. It is slightly different from the case in $4D$, as
we shall show, since there is no constraint like $\pi_1$, the structure
of constraints is much simpler. We find that there are six first
class constraints ($H, H_i, \Pi^{a0}$), $a=0,1,2,\ i=1,2$, together
with four second class constraints ($\Gamma^{1}, \Gamma^{2},
\Gamma^{12}, \pi$) in all, thus the degrees of freedom are two.

Following the procedure developed in Set.III, we can derive all the
ten constraints ($H_0, H_i, \Pi^{a0}, \Gamma^{1}, \Gamma^{2},
\Gamma^{12}, \pi$), where ``$a$" runs from 0 to 2 and $i=1,2$. The
Poisson brackets between those constraints are the same as those in
$4D$, see eqs.~(\ref{constraints1})-(\ref{constraints4}). Similarly,
when we define
\begin{eqnarray}
y_i=\{ H_0, \Gamma^i \},\qquad y_3=\{ H_0, \Gamma^{12} \},\label{3Dy}\\
x_0= \{ H_0, \pi \},\qquad x_i=\{\Gamma^i, \pi \},\qquad
x_3=\{\Gamma^{12},\pi \},\label{3Dx}
\end{eqnarray}
\begin{eqnarray}
A_i&=&\{\Gamma^i, \Gamma^{12} \}\nonumber\\
 &\approx& 2e[g^{0i}(g^{01}g^{2m} - g^{02}g^{1m}) + g^{1i} (g^{0m}g^{02} - g^{2m}g^{00})
- g^{i2}(g^{0m}g^{01} - g^{1m}g^{00})]\partial_m \phi,\label{3DA}
\end{eqnarray}
we can rewrite the self-consistent equations eq.~(\ref{equ}) in a
compact form
\begin{eqnarray}\label{3De}
M_{3D} \Lambda_{3D}=0,
\end{eqnarray}
where $\Lambda_{3D}=(1, \lambda_1, \lambda_2, \lambda_3, \lambda)^T$
and $M_{3D}$ is define by
\begin{equation}
M_{3D}=\left(
\begin{array}{ccccc}
 0 & y_1 & y_2 & y_3 & x_0\\
 -y_1 & 0 & 0 & A_1 & x_1\\
 -y_2 & 0 & 0 & A_2 & x_2\\
 -y_3 & -A_1 & -A_2 & 0 & x_3\\
 -x_0 & -x_1 & -x_2 & -x_3 & 0
\end{array}
\right).
\end{equation}
Since eq.~(\ref{3De}) has one non-zero solution $\Lambda_{3D}$, the
determinant of $M_{3D}$ should vanish, which is satisfied
automatically because $M_{3D}$ is a $5 \times 5$ antisymmetric
matrix. Thus, unlike the case in $4D$, there is no further
constraint for $f(T)$ gravity in $3D$. The rank of $M_{3D}$ is four,
which means that there are four independent equations for the four
Lagrange multipliers. We can derive all the Lagrange multipliers as
follow:
\begin{eqnarray}\label{multipliers}
\lambda_1&=&\frac{A_2 x_0+x_3 y_2-x_2 y_3}{A_1 x_2-A_2 x_1},\\
\lambda_2&=&\frac{-A_1 x_0-x_3 y_1+x_1 y_3}{A_1 x_2-A_2 x_1},\\
\lambda_3&=&\frac{y_1 x_2-y_2 x_1}{A_1 x_2-A_2 x_1},\\
\lambda &=&\frac{A_1y_2-A_2 y_1}{A_1 x_2-A_2 x_1}.
\end{eqnarray}
Since we are interested in the most general case, we require $A_1
x_2-A_2 x_1\neq 0$ here. Now it is clear that $\Gamma^{1},
\Gamma^{2}, \Gamma^{12}, \pi$ are second class constraints while $H,
H_i, \Pi^{a0}$ (Note that it is $H$, not $H_0$) are first class
constraints. Therefore, the degrees of freedom is two for $f(T)$
gravity in $3D$.

The discussions in this section and the above section can be
extended to the case in dimensions higher than two. In general,
there are $\frac{D(D-3)}{2}+D-1$ degrees of freedom for $f(T)$
gravity in $D$ dimensions. Since the calculations are very
complicated, we only show some key points here. Firstly, one should
note that the rank of the $\frac{D(D-1)}{2}\times \frac{D(D-1)}{2}$
matric eq.~(\ref{constraints1}) is $2(D-2)$. Thus, from the second
equation of eq.~(\ref{equ}) one can determine the Lagrange
multiplier $\lambda$ and derive $\frac{D(D-1)}{2}-2(D-2)-1$
secondary constraints. Secondly, note that those secondary
constraints contains the constraint like $\pi_1=\sqrt{\mid M\mid}$
eq.~(\ref{pi}). One can check that in $4$ dimensions the constraint
derived from the second equation of eq.~(\ref{equ}) and square root
of the determinant of $M$ eq.~(\ref{pi}) are exactly the same. So,
after imposing those secondary constraints and substituting
$\lambda$ into eq.~(\ref{equ}), there are $2(D-2)+1$ instead of
$2(D-2)+2$ independent equations for $\frac{D(D-1)}{2}$ Lagrange
multipliers $\lambda_{ab}$. Thirdly, the consistency condition of
those secondary constraints lead to $\frac{D(D-1)}{2}-2(D-2)-1$ more
equations for Lagrange multipliers $\lambda_{ab}$. Thus, all the all
the Lagrange multipliers can be determined and there are no further
secondary constraints. Finally, there are $2D$ first class
constraints and $D(D-1)-2(D-2)$ second class constraints, so the
degree of freedoms are $\frac{D(D-3)}{2}+D-1$ for $f(T)$ gravity in
$D$ dimensions which implies that the $D-1$ extra degrees are one
massive vector field or one massless vector field with one scalar
field. Incidentally, there are no independent degrees of freedom for
$f(T)$ gravity in $2D$ (Since $T=0$ and
$\Sigma_{\mu\nu\rho}=0$ in $2D$, it should be mentioned that Ferraro
and Fiorini also observed [22] the lack of dynamics in $2D$ $f(T)$ gravity.).

\section{Conclusions}

 We have established the Hamiltonian formulations of $f(T)$ gravity.
 In $4D$, we find that the six first class constraints corresponding to the local
lorentz invariance in $TG$ become second class constraints in $f(T)$
gravity, which implies that there are three extra degrees of freedom
and the local lorentz invariance is broken completely. In $3D$, the
constraint structure is much simpler and the independent degrees of
freedom are two. In addition, there are $D-1$ extra degrees of
freedom for $f(T)$ gravity in $D$ dimensions which implies that the
extra degrees of freedom correspond to one massive vector field or
one massless vector field with one scalar field.
From the conformal rescaling of the action eq.~(2.11), we observe that the vector degrees of freedom might
emerge from some kind of Higgs mechanism.
This problem needs further study in the future. We hope our results
will give some guidance for the cosmological perturbations of $f(T)$
gravity, where the gauge conditions and extra degrees of freedom
should carefully be treated when compared with Einstein gravity.
Besides, the
extra degrees of freedom are expected to play the role of dark
energy in $f(T)$ gravity. More studies on the
properties of the extra degrees of freedom are needed, such as whether the
extra degrees of freedom correspond to vector fields, and whether the fields are
stable, and so on, based on which one may obtain more insights into the
behaviors of dark energy in $f(T)$ gravity.

\section*{Acknowledgements}
R-X.M. would like to thank T. Wang, B. Li and T. P. Sotiriou for
email communications. M.L. and R-X.M. are supported by the NSFC
grants No.10535060, No.10975172, and No.10821504, and by the 973
program grant No.2007CB815401 of the Ministry of Science and
Technology  of China. Y-G.M. is supported by the NSFC grant
No.10675061.

\section*{Appendix}

We shall prove that all the Lagrange multipliers can be
determined from eq.~(\ref{MLamma}) and eq.~(\ref{C1}), thus there is
no further secondary constraint. The exact result of eq.~(\ref{C1})
is highly complicated due to eq.~(\ref{pi}), however, it is not
necessary to derive eq.~(\ref{C1}) exactly. Let us adopt  a technique of calculation to
overcome this difficulty.

Recall the results obtained in Set.IV. From eq.~(\ref{MLamma}), we
can derive one secondary constraint $\pi_1$ and obtain six
independent equations for the seven Lagrange multipliers. Without the
loss of generality, we can express all the other Lagrange
multipliers in terms of $\lambda_5$. Substituting the Lagrange
multipliers into eq.~(\ref{C1}),
\begin{eqnarray}\label{C11}
\{ \pi_1, H \}&=&\{ \pi_1, H_0 \}+\{ \pi_1, \Gamma^{cd}
\}\lambda_{cd} +\{ \pi_1, \pi \}\lambda \approx 0,
\end{eqnarray}
we can get one equation for $\lambda_5$ in the general form
\begin{eqnarray}\label{C12}
C=(B+E^i\partial_i)\lambda_5
\end{eqnarray}
Only under the very strict conditions $B=E^i=0$ can we derive
another secondary constraint $C=0$, otherwise we can solve
$\lambda_5$ from the above equation. Note that only the terms
$\{\pi_1, \Gamma^{cd} \}$ contribute to $E^i$. In view of
$\pi_1=C^{(mnl)}\partial_m \phi\partial_n \phi\partial_l \phi$,
where $C^{(mnl)}$ is independent of the terms such as $\partial_j
\Pi^{ak}$, $\partial_i T_{ajk}$ and $\partial_k \phi$, and it is
linear to the product of $\Pi^{ai}$ and $T_{aij}$, we find that
$E^i$ must take the form of $E^{imnl}\partial_m \phi\partial_n
\phi\partial_l \phi$, where $E^{imnl}$ contains no derivatives of
$\Pi^{ak}$, $T_{ajk}$ and $\phi$. If $E^i$ does not vanish
automatically, we cannot make it vanish by imposing the constraints
$H_0$, $H_i$, $\Gamma^{ab}$, and $\pi_1$. Since $H_0$ and $H_i$
contain the derivative of $\Pi^{ak}$, they are independent of $E^i$.
Because $\Gamma^{ab}$ contains no terms related to $\partial_m \phi$
it is also independent of $E^i$ (Note that though $\partial_m
\Gamma^{ab}$ contains the derivative of $\phi$, it also contains the
derivatives of $\Pi^{ak}$ and $T_{ajk}$. Thus, we cannot use
functions constructed from $\Gamma^{ab}$ and its derivatives to
eliminate $E^i$). Only $\pi_1$ can be used to eliminate $E^i$, but
there is only one $\pi_1$ which cannot eliminate all $E^i$. We
shall show below that $E^i$ is non-vanishing, therefore there is no
further secondary constraint and we can derive all the Lagrange
multipliers.

Suppose that we have derived $E^i=E^{imnl}\partial_m \phi\partial_n
\phi\partial_l \phi$ and $B$ strictly by applying the complicated
expression of $\pi_1$ eq.~(\ref{pi}). Then, imposing the condition
that $g_{\mu\nu}$ is diagonal, we can simplify $E^i$ , $B$ and
obtain $A'^i$ , $B'$. If $E'^i$ is non-vanishing, so is $E^i$. In
order to derive $E'^i$, we do not need to use the exact form of
$\pi_1$. In fact, eq.~(\ref{pi1}) is enough. Consequently,  our
technique is to apply eq.~(\ref{pi1}) to derive $E'^i$. It should be
stressed that if we want to derive the correct $B'$, we must
preserve the first order of the non-diagonal parts of $g_{\mu\nu}$
for $\pi_1$. However, for the derivation of $E'^i$, the zero order
of non-diagonal parts of $g_{\mu\nu}$ is enough.  Note that the
condition $g_{\mu\nu}$ is diagonal is not a gauge, but our technique
of calculation. We impose only this condition at the end of the
calculations to simplify the results.

Under the condition that $g_{uv}$ is diagonal, we can derive
\begin{eqnarray}\label{Ai}
E'^i=-h_5^i+\frac{A_{12}}{A_{11}}h_4^i+\frac{A_{32}}{A_{33}}h_6^i-\left(-h_2^i+\frac{A_{21}}{A_{11}}h_1^i
+\frac{A_{23}}{A_{33}}h_3^i\right)\frac{-x_5+x_4\frac{A_{12}}{A_{11}}+x_6\frac{A_{32}}{A_{33}}}{-x_2+x_1\frac{A_{21}}{A_{11}}+x_3\frac{A_{23}}{A_{33}}},
\end{eqnarray}
where $h_{*}^{\ i}$ is defined by
\begin{eqnarray}\label{hi}
\{ \Gamma^j(x), \pi_1 (y)\}&=&-h_j^{\
i}(x)\partial_{x_i}\delta(x-y)+\cdots,\nonumber\\
\{ \Gamma^{12}(x), \pi_1 (y)\}&=&-h_4^{\
i}(x)\partial_{x_i}\delta(x-y)+\cdots,\nonumber\\
\{ \Gamma^{13}(x), \pi_1 (y)\}&=&-h_5^{\
i}(x)\partial_{x_i}\delta(x-y)+\cdots,\nonumber\\
\{ \Gamma^{23}(x), \pi_1 (y)\}&=&-h_6^{\
i}(x)\partial_{x_i}\delta(x-y)+\cdots,
\end{eqnarray}
where the above `$\cdots$' stand for the terms without the
derivative of the delta function. Substituting $\pi_1$
eq.~(\ref{pi1}) into the above equations, we can derive all the
$h_{*}^{\ i}$. After some tedious calculations, we find that $E^i$
eq.~(\ref{Ai}) does not vanish even imposing all of the constraints.
As a result, we can solve all the Lagrange multipliers and there are
no further secondary constraints.

 \end{document}